\documentstyle[11pt,AATS,epsf]{article}	

\begin{document}	

\title{$JHK$ Spectroscopy of the $z=2.39$ Radio Galaxy 53W002 and Its
Companions
} 

\author{K. Motohara}
\affil{Subaru Telescope, National Astronomical Observatory of
 Japan, 650 North A'ohoku Place, Hilo, HI 96720, USA}
\author{T. Yamada}
\affil{National Astronomical Observatory of Japan, Mitaka, Tokyo 181-8588, Japan}
\author{F. Iwamuro, T. Maihara}
\affil{Kyoto University, Kitashirakawa, Kyoto 606-8502, Japan}


\begin{abstract}

We present low-resolution, near-IR $JHK$ spectra of
the weak z=2.39 radio galaxy 53W002 and its companion objects \#18 and \#19, 
obtained with OHS/CISCO on the Subaru Telescope.  
They cover rest-frame wavelengths of 3400-7200\AA, and many rest-optical 
emission lines are detected.
Contributions to the broad-band flux from these emission lines are found
 to be very large, up to 40\% in the $H$ and $K'$-bands and 30\% in the
 $J$-band.

\end{abstract}


\section{Introduction}
Recent narrow-band imaging has revealed the existence of a cluster of
Ly$\alpha$ emitters around the $z=2.39$ radio galaxy 53W002
 (Pascarelle et al. 1996a,b; Pascarelle et al. 1998; Keel et al. 1999).
These emitters have sub-galactic sizes, and are
thought to be ``building blocks'' which will merge into a luminous
galaxy at the present epoch.

To investigate the rest-frame optical nature of these objects, 
we carried out low-dispersion $JHK$ spectroscopy of 53W002, Object \#18,
and \#19 using the newly commissioned instrument of
the Subaru telescope, OH-airglow suppression spectrograph (OHS; Iwamuro
et al. 2001) and Cooled Infrared Spectrograph and Camera for OHS (CISCO;
Motohara et al. 1998).

\section{Detection of Strong Rest-Optical Emission Lines}
The resulting spectra of these objects are shown in Figure 1.
Both 53W002 and Object \#18 show very strong [O\,{\sc
iii}] and H$\alpha$+[N {\sc ii}] lines.
The Balmer jump is also detected in the continuum of 53W002.
Object \#19, which is known to be a quasar, shows a power-law continuum and
a broad (8000 km s$^{-1}$),
strong ($3\times10^{43}$ erg s$^{-1}$; $H_0=65$ km s$^{-1}$ Mpc$^{-1}$,
$q_0=0.1$) H$\alpha$ line. 
The contribution of these lines to the broad-band flux is as high as 
45\%, as shown in Table 1.

\begin{figure}[t]	
\plotfiddle{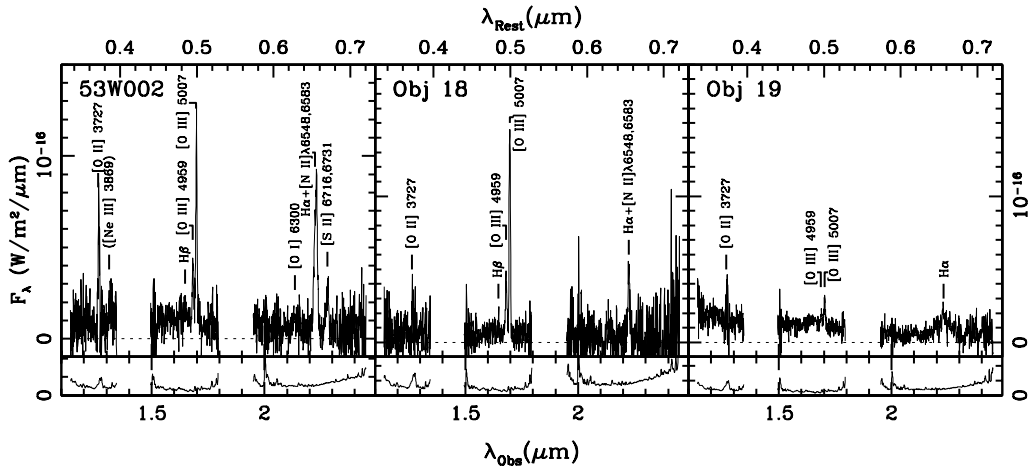}{5.5cm}{0}{75}{75}{-240}{-130}	
\caption{The $JHK$ spectra. Lower box shows the 1$\sigma$ noise
 level calculated from the background level.}
\end{figure}

\begin{table}[b]
\begin{center}		
\caption{Contribution of the emission lines to the broad-band flux. }

\begin{tabular}{lccc}  	
\tableline		
Object & \makebox[13mm]{$K^{\prime}$} &  \makebox[13mm]{$H$} &  \makebox[13mm]{$J$} \\
\tableline		
53W002 & \hspace*{3.1mm}42\% & 30 & 22\\
Object \#18 & 32 & 45 & 11\\
Object \#19 & 29 & 6 & 4\\
\tableline \tableline
\end{tabular}
\end{center}
\end{table}

\section{Emission Line Diagnosis}
Because the wavelength resolution is low
($\lambda/\Delta\lambda= 200-400$), we deconvolved the
blended H$\alpha$+[N {\sc ii}] lines by multiple Gaussian
fitting. We then estimated the dust extinction from H$\alpha$/H$\beta$
ratio, assuming SMC extinction curve.
The estimate of $E(B-V)$ is 0.14 for 53W002 and 0.50 for Object \#18.

In Figure 2, two diagrams of reddening corrected emission-line ratios are
presented. 
One show [N\,{\sc ii}]$\lambda$6583/H$\alpha$ versus [O\,{\sc
iii}]$\lambda$5007/H$\beta$ and the other [O\,{\sc
ii}]$\lambda$3727/[O\,{\sc iii}]$\lambda$5007 versus [O\,{\sc
iii}]$\lambda$5007/H$\beta$. 

We next carried out photoionization calculation using CLOUDY94 (Ferland
 2000), and over-plotted the results on Figure 2.
These results show that the emission line ratios of 53W002 are well
reproduced by a cloud with electron density $n_{\rm
e}=1\times10^{3-4}\,\rm cm^{-3}$ and solar metallicity, ionized by an
$\alpha=-0.7$ power-law continuum.
Object \#18 seems to be not a star-forming galaxy but a type 2 AGN, and 
its line ratios are reproduced by a cloud of solar metallicity,
ionized by $\alpha=-1.5$ power-law continuum.

Both 53W002 and Object \#18 show high metallicity (solar abundances).
We suggest that they are produced by starburst activity during 
merger events with surrounding objects, for which we find evidence in
our spectrum of 53W002 in the form of the Balmer discontinuity at 4000\AA.

\begin{figure}[t]	
\plotfiddle{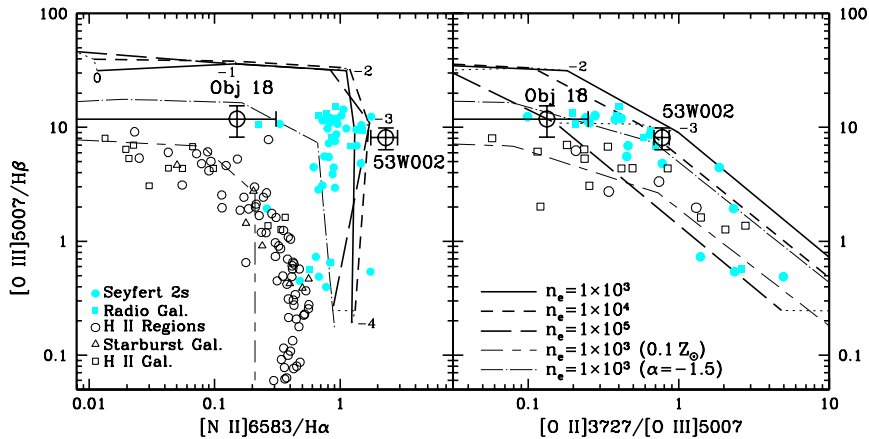}{5.5cm}{0}{60}{60}{-180}{-100}	
\caption{Reddening corrected line ratios of 53W002 and Object \#18 (open
 circles with error bars). Spectral index is assumed to be $\alpha=-0.7$ except for one case of
 $\alpha=-1.5$ shown by thin dot-dashed line.
 The metallicity is set to be $Z=1.0 Z_{\sun}$, except for one case
 of $Z=0.1 Z_{\sun}$ shown by thin dash-long-dashed line.
 Ionization parameter varies along each curve, and 
 representative points are labeled by their powers.
}
\end{figure}


\begin{references}
\reference Ferland, G.J.\ 2000 RMxAC, 9, 153F
\reference Iwamuro, F., Motohara, K., Maihara, T., Hata, R., Harashima, T., \& Sekiguchi, K.\ 2001, \pasj, in press.
\reference Keel, W.C., Cohen, S.H., Windhorst, R.A., \& Waddington, I.\
 1999, \aj, 118, 2547
\reference Motohara, K., Maihara, T., Iwamuro, F., Oya, S., Imanishi,
 M., Terada, H., Goto, M., Iwai, J., Tanabe, H., Tsukamoto, H, \&
 Sekiguchi, K.\ 1998, Proc.\ SPIE, 3354, 659 
\reference Pascarelle, S.M., Windhorst, R.A., Driver, S.P., Ostrander,
 E.J., \& Keel, W.C.\ 1996a, \apj, 456, L21
\reference Pascarelle S.M., Windhorst R.A., Keel W.C., \& Odewahn S.C.\
 1996b, Nature, 383, 45
\reference Pascarelle, S.M., Windhorst, R.A., \& Keel, W.C.\ 1998, \aj, 116,
 2659
\end{references}
\end{document}